\documentclass[preprint]{aastex}

\shorttitle{Predicting Coronal Mass Ejections Using Machine Learning Methods}
\shortauthors{Bobra and Ilonidis}

\usepackage{natbib}
\usepackage{hyperref}
\bibpunct{(}{)}{;}{}{,}{,}

\begin{document}
\title{Predicting Coronal Mass Ejections Using Machine Learning Methods}
\author{M. G. Bobra and S. Ilonidis}
\affil{W.W. Hansen Experimental Physics Laboratory, Stanford University, Stanford, CA 94305}

\begin{abstract}
\noindent Of all the activity observed on the Sun, two of the most energetic events are flares and Coronal Mass Ejections (CMEs). Usually, solar active regions that produce large flares will also produce a CME, but this is not always true \citep{yashiro05}. Despite advances in numerical modeling, it is still unclear which circumstances will produce a CME \citep{webb12}. Therefore, it is worthwhile to empirically determine which features distinguish flares associated with CMEs from flares that are not. At this time, no extensive study has used physically meaningful features of active regions to distinguish between these two populations. As such, we attempt to do so by using features derived from [1] photospheric vector magnetic field data taken by the Solar Dynamics Observatory's Helioseismic and Magnetic Imager instrument and [2] X-ray flux data from the Geostationary Operational Environmental Satellite's X-ray Flux instrument. We build a catalog of active regions that either produced both a flare and a CME (the positive class) or simply a flare (the negative class). We then use machine-learning algorithms to [1] determine which features distinguish these two populations, and [2] forecast whether an active region that produces an M- or X-class flare will also produce a CME. We compute the True Skill Statistic, a forecast verification metric, and find that it is a relatively high value of $\sim$0.8$\pm$0.2. We conclude that a combination of six parameters, which are all {\it intensive} in nature, will capture most of the relevant information contained in the photospheric magnetic field.
\pagebreak
\end{abstract}

\section{Introduction}
\label{section:intro}

A Coronal Mass Ejection (CME) is a rapid ejection of magnetic flux and plasma from the Sun into interplanetary space \citep{forbes00}. They are typically observed in coronagraphs, instruments that block out the solar disk to observe the corona, as an arc of bright light streaking through space. Over the last couple decades, there has been a general consensus that CMEs and flares are in fact related as part of ``a single magnetically-driven event" \citep{webb12}, wherein a flare unassociated with a CME is often referred to as a {\it confined} or {\it compact} flare. In general, the more energetic a flare, the more likely it is to be associated with a CME \citep{yashiro05} --- but this is not, by any means, a rule. For example, \citet{sun15} found that the largest active region in the last 24 years, which produced 6 X-class flares, did not produce a single CME. 

It is unclear why some solar activity triggers both a flare and a CME and other activity simply triggers a confined flare. Though many numerical models describe the physical processes by which flares and CMEs are produced, outstanding questions remain about the magnetic topologies of these eruptive events and how they release so much energy so fast \citep{webb12}. In the absence of a definitive physical theory, it is worthwhile to empirically determine which features distinguish flares associated with CMEs from flares that are not. 

One way to distinguish between the two is by use of a machine-learning algorithm. One specific class of machine-learning algorithms, known as supervised binary classifiers, will, given a set of features characterizing an event, predict which of two classes an event belongs to. In other words, the classifier learns from some data for which it already knows the answer, and applies this knowledge to predict an answer for other, similar data. One such algorithm is called a Support Vector Machine (SVM), which is advantageous for its ability to predict which class an event falls into even in cases where the number of features is large. SVMs can also identify non-linear relationships between features (by using kernels to map these features into a higher dimensional space) and thus may potentially unearth previously unknown properties in the data. 

Various studies report correlations between features of solar active regions and CME productivity. \citet{qahwaji08} used a non-linear SVM to predict whether or not a flare will initiate a CME. However, their study only used the duration of these events, rather than physically meaningful quantities, to distinguish between the populations. \citet{falconer08} report correlations that are linear within their feature space between physically meaningful parameterizations of the line-of-sight component of the magnetic field, as measured by the Michelson Doppler Imager (MDI) on the Solar and Heliospheric Observatory (SoHO), near active region neutral lines and used these to predict CMEs with a $75\%$ success rate. \citet{falconer06} reached a similar conclusion using slightly different parameterizations of active region neutral lines and vector magnetic field data from the Marshall Space Flight Center vector magnetograph. However, these studies limit their exploration to solely CME-productive active regions. As far as we can tell, no study has used physically meaningful features derived from the photospheric magnetic field to distinguish between flares that are CME productive and those that are not.

To this end, we use a [1] SVM and features derived from photospheric vector magnetic field data taken by the Solar Dynamics Observatory's (SDO) Helioseismic and Magnetic Imager (HMI) instrument to forecast whether an active region that produces an M1.0-class flare or higher will also produce a CME, and [2] feature selection algorithm to determine which features distinguish these two populations. This paper is organized as follows: in Section \ref{section:Data}, we describe the solar data and the event catalog; in Section \ref{section:Algorithms}, we describe the application of the machine-learning algorithms to these data; and, in Section \ref{section:Results}, we present our results. 

The code used to conduct this study is publicly and permanently available in the Stanford Digital Repository\footnote{\url{https://purl.stanford.edu/wt605kh4712}}; the data used to conduct this study are all public and their repositories are detailed in Section \ref{section:Data}. Both the feature selection and SVM implementations are from the python scikit-learn library \citep{pedregosa11}, which is open-source, widely-used, and well-established.
 
\section{Data}
\label{section:Data}

\begin{deluxetable}{lllccc}
\tabletypesize{\scriptsize}
\rotate
\tablecaption{Features and F-scores. \label{tbl1}}
\tablewidth{0pt}
\tablehead{\colhead{Keyword} & \colhead{Description} &  \colhead{Formula} & \colhead{Scaling} & \multicolumn{2}{c}{Feature Ranking} \\
\colhead{} & \colhead{} &  \colhead{} & \colhead{} & \colhead{24 hours} &\colhead{48 hours}}
\startdata
{\sc meangbh} & Mean gradient of horizontal field & $\overline{\left|{\nabla B_h}\right|} = \frac{1}{N} \sum \sqrt{\left(\frac{\partial B_h}{\partial x}\right)^2 + \left(\frac{\partial B_h}{\partial y}\right)^2}$ & I & 1 & 13 \\
{\sc meanjzh} & Mean current helicity ($B_{z}$ contribution) & $\overline{H_c} \propto \frac{1}{N} \sum B_z \cdot J_z $  & I & 2 & 2 \\
{\sc meanalp} & Mean characteristic twist parameter, $\alpha$ & $ \alpha_{total} \propto \frac{\sum J{_z} \cdot B_z}{\sum B{^2_z}} $  & I & 3 & 5 \\
{\sc meangbt} & Mean gradient of total field & $\overline{\left|{\nabla B_{\rm tot}}\right|} = \frac{1}{N} \sum \sqrt{\left(\frac{\partial B}{\partial x}\right)^2 + \left(\frac{\partial B}{\partial y}\right)^2}$  & I & 4 & 18 \\
{\sc meanpot} & Mean photospheric magnetic free energy & $ \overline{\rho} \propto \frac{1}{N} \sum \left( \vec{\textit{\textbf B}}^{\rm Obs} - \vec{\textit{\textbf B}}^{\rm Pot} \right)^2 $  & I & 5 & 12 \\
{\sc meanshr} & Mean shear angle  & $ \overline{\Gamma} = \frac{1}{N} \sum \arccos \left( \frac{\vec{\textit{\textbf B}}^{\rm Obs} \cdot \vec{\textit{\textbf B}}^{\rm Pot}}{|B^{\rm Obs}|\,|B^{\rm Pot}|} \right)$   & I & 6 & 9 \\
{\sc shrgt45} & Fraction of Area with Shear $> 45^\circ$  & Area with Shear $>45^\circ$ / Total Area  & I & 7 & 6\\
{\sc totpot} & Total photospheric magnetic free energy density & $ \rho_{tot} \propto  \sum \left( \vec{\textit{\textbf B}}^{\rm Obs} - \vec{\textit{\textbf B}}^{\rm Pot} \right)^2 dA $  & E & 8 &11 \\
{\sc meanjzd} & Mean vertical current density & $\overline{J_z} \propto \frac{1}{N} \sum \left(\frac{\partial B_y}{\partial x} - \frac{\partial B_x}{\partial y}\right) $  & I & 9 & 10 \\
{\sc usflux} & Total unsigned flux & $\Phi = \sum|B_{z}|dA$  & E & 10 & 3 \\
{\sc meangam} & Mean angle of field from radial &  $\overline{\gamma} = \frac{1}{N} \sum \arctan\left(\frac{B_h}{B_z}\right)$ & I & 11 & 14 \\
{\sc totusjz} & Total unsigned vertical current & ${J_{z_{total}}} =  \sum |J_{z}|dA$   & E & 12 & 7 \\
{\sc absnjzh} & Absolute value of the net current helicity & ${H_{c_{abs}}} \propto \left| \sum B_z \cdot J_z \right|$  & E & 13 & 1 \\
{\sc area\_acr} &  Area of strong field pixels in the active region &  Area $ = \sum$ Pixels  & E & 14 & 15 \\
{\sc r\_value} &  Sum of flux near polarity inversion line &  $\Phi = \sum|B_{LoS}|dA$ within R mask  & E & 15 & 17 \\
{\sc totusjh} & Total unsigned current helicity  & ${H_{c_{total}}} \propto \sum |B_z \cdot J_z|$  & E & 16 & 4 \\
{\sc } & Flare Class & $FC = CM $ & neither & 17 & 19 \\
{\sc savncpp} & Sum of the modulus of the net current per polarity & $J_{z_{sum}} \propto \Big\vert \displaystyle\sum\limits^{B{_z^+}} J{_z}dA \Big\vert + \Big\vert \displaystyle\sum\limits^{B{_z^-}} J{_z}dA \Big\vert $  & E & 18 & 8 \\
{\sc meangbz} & Mean gradient of vertical field & $\overline{\left|{\nabla B_z}\right|} = \frac{1}{N} \sum \sqrt{\left(\frac{\partial B_z}{\partial x}\right)^2 + \left(\frac{\partial B_z}{\partial y}\right)^2}$  & I & 19 & 16 \\
\enddata
\tablecomments{The {\it Keyword} column indicates the name of the FITS header keyword in the SHARP data series. The {\it Scaling} column indicates whether the parameter in question scales in an extensive (E) or intensive (I) manner per \cite{welsch09}. One feature, called the Flare Class, is not available as a SHARP FITS header keyword; rather, this is computed from the GOES data. The flare class is similar in concept to the flare index, described in \citet{antalova96}, except it is calculated for one flare. We define the flare class, $FC$ as follows:  $FC=CM$, where $C$ is a constant that equals 10.0 in the case of X-class flares and 1.0 in the case of M-class flares and $M$ is the magnitude of the flare.}
\end{deluxetable}


In May 2010, the SDO HMI instrument began producing full-disk vector magnetic field data every 720 seconds. It is the first space-based instrument to continuously measure the full-disk photospheric vector magnetic field \citep{schou12}. In 2014, the HMI team released a data product, called Space-weather HMI Active Region Patches (SHARPs), which is publicly available at the Joint Science Operations Center (JSOC)\footnote{\url{http://jsoc.stanford.edu}}. The SHARP data contains a vector magnetic field map of every single solar active region observed since SDO's launch, each of which are automatically tracked for its entire lifetime, and 18 keywords that parameterize the vector magnetic field within these regions \citep{bobra14}. 

The SHARP parameterizations include measurements of physical quantities, such as the magnetic flux contained in an active region; they also include proxies of physical quantities, such as the current helicity, which cannot be measured in its entirety within a single plane. Further details about the definition of and motivation behind each parameter can be found in \citet{bobra14}. A list of all 18 SHARP features is included in Table \ref{tbl1}. Between 2010 May 1 and 2015 May 1, the HMI data processing pipeline automatically detected 3023 active regions and computed 18 parameterizations of each active region at every 720-second interval throughout its lifetime, resulting in $\sim$36 million unique data points.

We use data from the SoHO Large Angle and Spectrometric Coronagraph Experiment (LASCO) instrument and both coronagraphs on the Solar Terrestrial Relations Observatory (STEREO) Sun Earth Connection Coronal and Heliospheric Investigation (SECCHI) instruments to determine whether or not an active region produced a CME. These data are compiled into a NASA Space Weather Research Center database called Space Weather Database Of Notification, Knowledge, Information (DONKI)\footnote{\url{http://kauai.ccmc.gsfc.nasa.gov/DONKI/}}. This database contains a table of solar flares and, within this table, a column indicating whether or not any given flare produced a CME. We query this database for events between 2010 May 1 and 2015 May 1 to build our positive class; as such, the positive class is defined as events for which a flaring active region also produced a CME. We do not consider flares below the M1.0-class, as they release a limited amount of energy and thus are unlikely to disturb the near-Earth environment. Some flares in this database are not associated with an active region. As it is not possible to compute features for events unassociated with an active region, we reject these flares from our sample. 

We use data from the X-ray Flux instrument aboard the Geostationary Operational Environmental Satellite (GOES) to determine whether or not an active region produced an X- or M-class flare. The GOES flare list can be queried from the Heliophysics Events Knowledgebase\footnote{\url{http://www.lmsal.com/hek}} \citep{hurlburt12} using a python library called SunPy\footnote{\url{http://docs.sunpy.org/en/stable/code\_ref/instr/goes.html}} \citep{sunpy}. We use this database to build our negative class by compiling a list of all the M1.0-class flares that occurred between 2010 May 1 and 2015 May 1 and then removing from this list the events in the positive class. As such, the negative class is defined as events for which a flaring active region did not produce a CME. We also include an additional feature from this database called the flare class. The formula for the flare class is described in Table \ref{tbl1}. Some flares in the GOES flare list are not associated with an active region. We reject these flares from our sample. We also reject all flares that are not within $\pm 70^{\circ}$ of central meridian during the GOES X-ray Flux peak time as the the signal-to-noise in the HMI vector magnetic field data decreases significantly beyond this longitude. 

We then compile all 18 SHARP active region features that characterize every event in both the positive and negative class at exactly $t$ hours before the GOES X-ray flare peak time, where $t$ ranges from from 4 to 48 hours in 4 hour intervals. In other words, we use features from exactly $t$ hours before an event to predict the event. These prediction times are somewhat arbitrary, but the $t$=24 and $t$=48 hour cases are common in the literature (e.g. \citealp{ahmed13}) and may be a useful lead time in an operational setting. We further restrict the number of events to those where the [1] absolute value of the radial velocity of SDO is less than 3500 m/s (see Section 7.1.2 of \citet{hoeksema14} for a discussion about the periodicity in magnetic field strength due to the orbital velocity of SDO) and [2] HMI data are of high quality (see Section Appendix A of \citet{hoeksema14} for a discussion about HMI data quality; we essentially throw out observables produced during bad conditions) at this time. As such, we are left with 56 events in the positive class and 364 events in the negative class for the time period between 2010 May 1 and 2015 May 1. 

In the case of predicting which flaring active regions produce a CME, we expect that the number of events in the positive class is much smaller than the number in the negative class. \citet{yashiro05} showed that while more than 80\% of X-class flares are associated with CMEs, this number drops as a power law with decreasing flare class. On average, about 60\% of M-class flares produce a CME; even so, there is a great disparity across the M-class, where M1.0-class to M1.8-class flares are only $\sim$44\% likely to produce a CME. As such, we have $\sim$6.5 times more events in the negative class than in the positive one: our negative class includes 364 events, 230 of which are within the M1-range, whereas our positive class includes 56, where 7 are within the M1-range.

\section{Algorithms}
\label{section:Algorithms}
\subsection{Feature Selection}
\label{subsection:FeatureSelection}
Some of the features within a dataset may be powerful in distinguishing between the positive and negative class, whereas others may be redundant or irrelevant. To identify features in the former category, we use a univariate feature selection method, which is implemented in the feature selection module of the scikit-learn library, for feature scoring. This involves computing an F-score, which is a ratio between two different measures of variance within the data (e.g. Section 4.3 of  of \citealp{rietveld05}). For two groups (i.e. the positive and negative class), the F-score for any given feature $i$ is given by:

\begin{equation}
F(i)=\frac{{n^+}(\bar{x}^+_i - \bar{x}_i)^2+{n^-}(\bar{x}^-_i - \bar{x}_i)^2}{\frac{1}{N-2}\left[\sum_{j=1}^{n^+}(x^+_{j,i} - \bar{x}^+_i)^2 + \sum_{j=1}^{n^-}(x^-_{j,i} - \bar{x}^-_i)^2 \right]}
\end{equation}

\noindent where $\bar{x}^+_i$ is the average of the values of feature $i$ over the positive-class examples, $\bar{x}^-_i$ is the average of the values over the negative-class examples, $\bar{x}_i$ is the average of the values over the entire dataset, $n^+$ is the total number of positives examples in the dataset, $n^-$ is the number of negative examples, and $N$ is the total number of examples. The numerator measures the variance between each class for a given feature, and the denominator measures the variance within each class for a given feature. If this ratio is small, both groups have a similar population mean. If this ratio is large, both groups have different population means. The univariate feature selection method ranks features according to their F-score. 

\subsection{Classification}
\label{subsection:Classification}

Our predictive model is based on a SVM, which has been used in many flare prediction studies (e.g. \citealp{li07}; \citealp{ahmed13}; \citealp{bobra15}). The SVM algorithm is a non-probabilistic binary classifier that determines whether the events in the positive and negative class are separable. More specifically, given a set of $N$ observations

\begin{equation}
x^{(i)}=\left(x^{(i)}_1, x^{(i)}_2, ..., x^{(i)}_d \right)
\end{equation}

\noindent with $i \in {1, 2, ..., N} $, where $d$ is an integer representing the number of features, and a corresponding set of labels $y^{(i)}= \pm 1$, where $+1$ is a positive example and $-1$ is a negative one, the SVM algorithm tries to find a hypersurface that separates the two classes. 

This can be accomplished by enlarging the original $d$-dimensional feature space $\vec{x}$, in which the decision boundary
between the two classes may be non-linear, into a higher dimensional feature space $\phi(\vec{x})$, in which the decision boundary between the two classes may be linear, i.e. the decision boundary is the hyperplane $\omega^T \phi(\vec{x}) + b = 0$. Here, $\phi(\vec{x})$ is the function that maps the original data into the higher dimensional space. One way to do this is by use of kernel functions $K\left(\vec{x}^{(i)},\vec{x}^{(j)} \right)$ defined as:

\begin{equation}
K\left(\vec{x}^{(i)},\vec{x}^{(j)} \right) = exp \left(-\gamma \sum_{k=1}^{d} \left(x^{(i)}_k - x^{(j)}_k \right) \right)
\end{equation}

\noindent where $\gamma$ is a positive tuning parameter that controls how far the influence of a single training observation extends to affect the classification of a new observation.

If the observations of the positive and negative class can be perfectly separated with a hyperplane, the SVM algorithm will find the maximal margin hyperplane, which is the hyperplane that has the farthest minimum distance to the observations in the dataset. However, if there exists no hyperplane that perfectly separates the positive class from the negative one, the SVM algorithm will allow a certain number of observations, called support vectors, to be on the wrong side of the margin or hyperplane. Each one of these observations is associated with a non-negative slack variable $\varepsilon_i$ that measures the degree of misclassification of the $i$th observation. If there are $N$ slack variables $\varepsilon_i$, the sum $\sum_{i=1}^{N}$ is the total error of the support vector observations. Generally, there is a tradeoff between the total error of the support vectors and the width of the margin between the positive and negative class. If the margin is wide, many observations violate the margin and the total error is too large. If, on the other hand, a solution includes only a small number of support vectors with low total error, the margin between the positive and negative class will be too narrow. This can be expressed as an optimization problem, where the vectors $\omega$ and $\varepsilon$ in the cost function, defined as:

\begin{equation}
\label{eqn1}
\mathrm{cost} = \left(\frac{1}{2} \omega^T \omega + C_1 \sum_{i=1}^{N^+} \varepsilon_i + C_2 \sum_{i=1}^{N^-} \varepsilon_i \right)
\end{equation}

\noindent are minimized according to the following constraint:

\begin{equation}
y^{(i)}\left(\omega^T \phi(\vec{x}^{(i)}) + b \right)\ge 1 -\varepsilon_i
\end{equation}

\noindent where $C_1$ and $C_2$ are non-negative tuning parameters for the positive and negative class, and $N^+$ and $N^-$ are the number of observations in the positive and negative class. The first term in Equation \ref{eqn1} is the inverse of the margin and the remaining terms are the total error of the support vectors. The SVM algorithm solves this optimization problem until the cost function of Equation \ref{eqn1} varies by less than a specified tolerance level. Once this optimization problem is solved, a new observation can be classified by determining on which side of the hyperplane it lies. 

In the case that $N^+ \simeq N^-$, $C_1$ can be set equal to $C_2$. If, as in our case, $N^+ \ll N^-$, the algorithm is more likely to favor the majority class. To favor both classes equally under these circumstances, we assign $C_1$ to be greater than $C_2$. In this way, the algorithm is receives a greater penalty for misclassifying minority class events than misclassifying majority class ones, and therefore does not neglect the minority class. 

\section{Results}
\label{section:Results}
\subsection{Feature Selection Results}
\label{subsection:FeatureSelectionResults}

Table \ref{tbl1} lists the ranking for each feature, for a prediction both 24 and 48 hours before the GOES X-ray flare peak time. The features are ordered such that the feature with the highest predictive power in the $t$=24 hour case, which is the mean gradient of the horizontal field, is listed first, and the remaining features are arranged in order of decreasing rank. The absolute value of the net current helicity has the highest predictive power for the $t$=48 hour case.

Of all of the features listed in Table \ref{tbl1}, the flare class, which we know from a priori constructing a dataset that only includes flaring active regions, is the only one that does not characterize an active region at a time $t$ before an event. Rather, the flare class can only be determined after an event. However, including the flare class in our feature selection algorithm allows us the opportunity to determine whether eruption potential (i.e. the probability of a flaring active region producing an associated CME) is simply a function of flare size. 

We find that the flare class is ranked 17 in the $t$=24 hour case and 19 in the $t$=48 hour case. We surmise that there are two reasons for such a result: [1] that our dataset is heavily biased by M-class flares (in the positive class, 73\% of our examples include M-class flares; this number is 98\% for the negative class), and [2] eruption potential is not solely a function of more energy within an active region system. Because the flare class is the only feature that cannot be used in an operational setting, we do not include the flare class in our predictive model.

We also find that some of the features change rank depending on the prediction time. For example, the feature which is highest ranked in the $t$=24 case ends up with a rank of 13 in the $t$=48 hour case. We find that this feature is the highest-performing for all prediction times less than 24 hours and is one of the lowest-performing for all prediction times between 24 and 48 hours. This may be a signature of different physical processes occurring within an active region over time.

Though the mean gradient of the horizontal field and the absolute value of the net current helicity are highest ranked for the $t$=24 and $t$=48 hour cases, respectively, this does not necessarily mean that they have any predictive power on their own. In order to determine which, if any, of the features can distinguish flaring active regions that produce CMEs from those that do not, we use a classification algorithm. If the classification algorithm performs well for any given feature or combination of features, the features are useful. Otherwise, the features are unable to distinguish between the two populations.

\subsection{Classification Results}
\label{subsection:ClassificationResults} 

\begin{figure}
\plotone{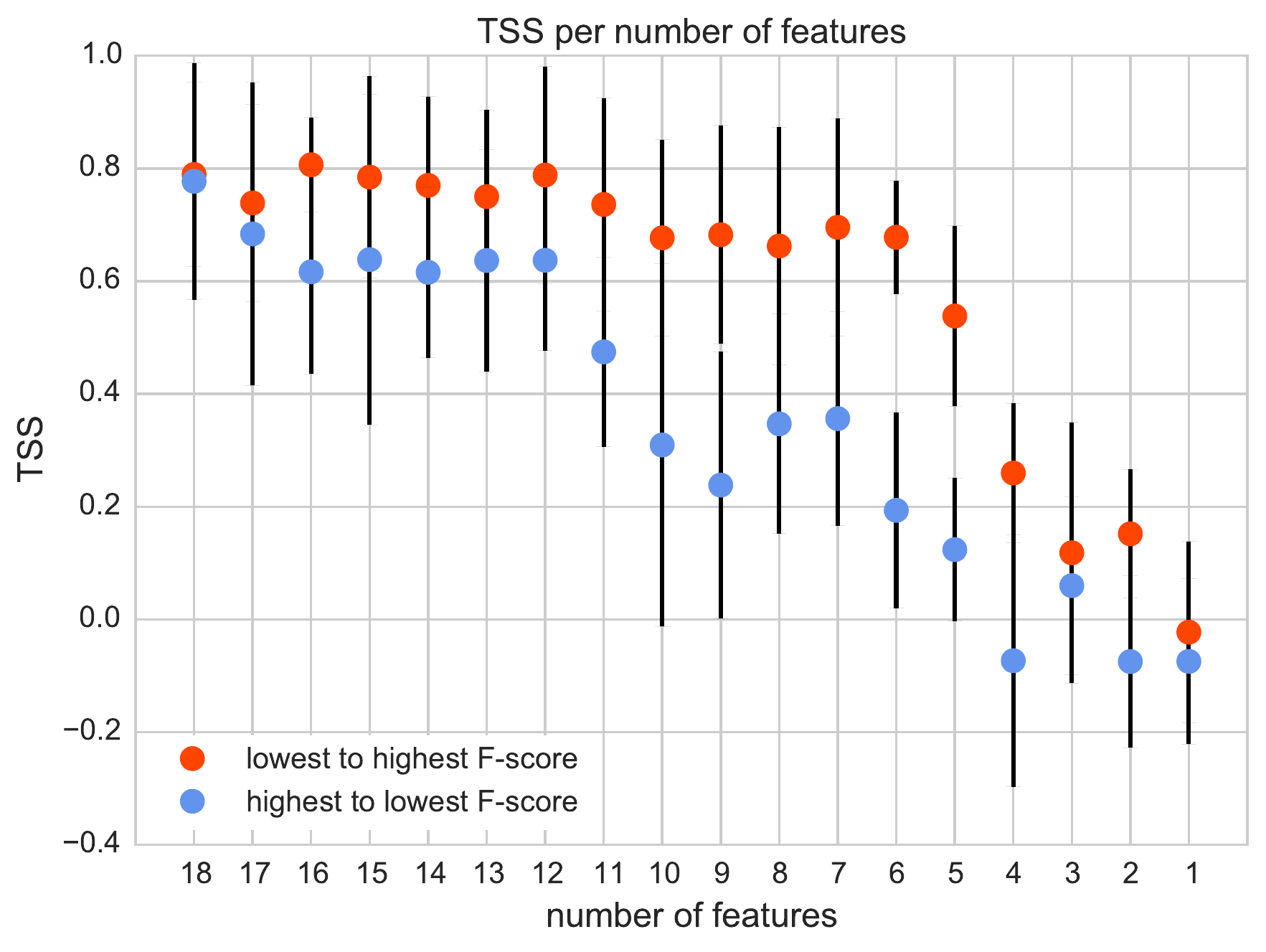}
\caption{TSS per number of SHARP features. For the red points, the features are arranged from lowest to highest F-score; for the blue points, this arrangement is reversed. These TSS values are computed for a prediction exactly 24 hours before the GOES X-ray flare peak time and with the parameters detailed in Section \ref{subsection:ClassificationResults}.}
\label{fig:TSS_features}
\end{figure}

\begin{figure}
\plotone{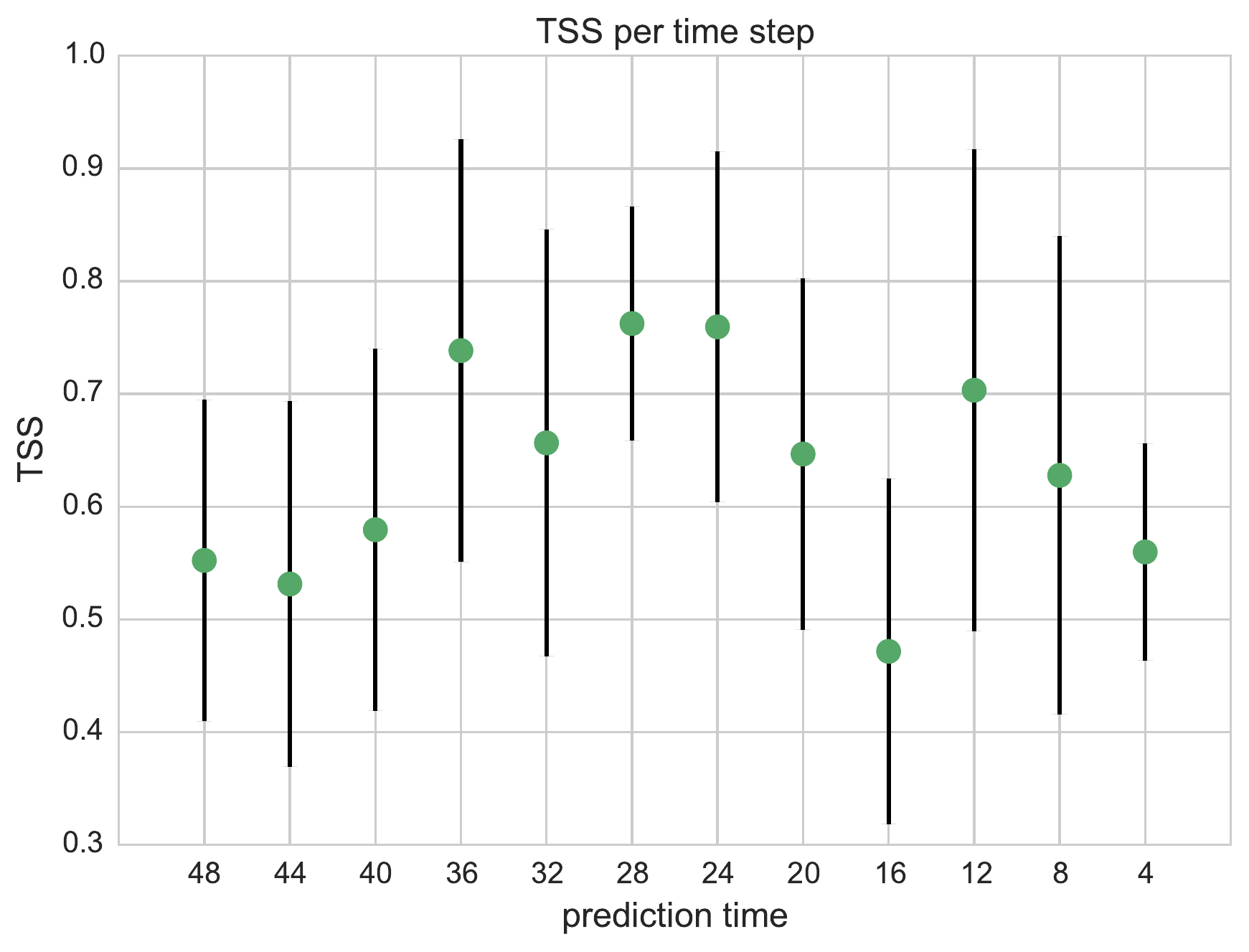}
\caption{TSS as a function of prediction time. These TSS values are computed using all 18 SHARP features and the parameters detailed in Section \ref{subsection:ClassificationResults}.}
\label{fig:TSS_time}
\end{figure}

We train the SVM on a subset of events, and then test it on the remaining events. In general, it is good practice to select only high-ranking features, as identified by the univariate feature selection algorithm, to train the SVM. However, this does not necessarily mean that the information contained in a low-ranking feature is absolutely useless. Features that are individually useless can, when combined, significantly improve the performance of a predictive model \citep{guyon03}. On the other hand, features that are absolutely useless can significantly worsen the performance of a predictive model by forcing the model to fit to noise. This process is called overfitting. In particular, overfitting becomes a concern when the number of features is large compared with the number of examples \citep{guyon03}. Since this is not the case here (we have 18 SHARP features and 420 examples), we use all 18 SHARP features to train the SVM (we do not use the flare class feature to train or test the SVM).

One way to test whether a set of features are absolutely versus individually useless is by computing a classification metric, which quantifies an algorithm's performance, on the testing set. Since the algorithm has no knowledge of the examples in the testing set, a model that has overfit the data in the training set will produce a low classification metric when applied to the data in the testing set. 

Many classification metrics exist in the literature; in the case of a highly imbalanced class ratio, some of these metrics are useful and some are not. For an in-depth review of the impact of class imbalance on classification metrics, see Section 4 of \citet{bobra15} and \citet{bloomfield12}. We prefer the True Skill Score (TSS) to all the other metrics as it is insensitive to the class imbalance ratio and thus best for comparison with other groups. The TSS is defined as follows:

\begin{equation}
\mathrm{TSS}=\frac{\mathrm{TP} \times \mathrm{TN}-\mathrm{FP} \times \mathrm{FN}}{\mathrm{P} \times \mathrm{N}}=\frac{\mathrm{TP}}{\mathrm{TP}+\mathrm{FN}}-\frac{\mathrm{FP}}{\mathrm{FP}+\mathrm{TN}} \label{TSS}
\end{equation} 

\noindent where TP is the number of true positives, FN the number of false negatives, TN the number of true negatives, and FP the number of false positives. The TSS is symmetrically distributed about 0: i.e., it ranges from [-1, 1] where 0 represents no skill and a negative value represents a perverse prediction. 

To train the SVM, we use the following parameters: $\gamma$ = 0.075, $C_1$ = 26.0, $C_2$ = 4.0. We determined these values through trial and error and tailored them to maximize the TSS. There are many different ways to separate a dataset into training and testing sets. A common way to do it is simply to divide the data into two groups and assign a smaller sample, e.g. 30\%, to the testing set, and then use the remaining sample for the testing set. However, since the positive sample size is quite small (both objectively and compared with the negative sample size) in our case, it is difficult to get an adequate number of events in both the training and testing sets. 

To solve this problem, we use the stratified $k$-folds cross-validation method, which makes $k$ partitions of the data set and uses $k$-1 folds for training and 1 fold for testing. The stratification preserves the ratio of positive to negative examples per fold. We then permute over the partitions such that each partition eventually makes its way into the testing set. For each individual testing set, we calculate the TSS. We report the average of the TSS over the total number of testing sets as the final TSS; we report the standard deviation of these TSS values as the error in the final TSS.

In order to use the stratified $k$-folds cross-validation method, we must first select a value of $k$. $k$ can be arbitrarily defined and take any value between 2 and the total number of examples. As $k$ approaches the total number of examples, the $k$-fold method reduces to the Leave One Out method, in which only one example is in the testing set and all other examples are in the training set. \citet{kohavi95} and other studies recommend the stratified 10-fold cross-validation to reduce variance and bias. Here, we test their recommendation by computing the TSS using 50 $k$ values, ranging from 2 to 52, and confirm that high $k$-values result in a high variance. As such, we use find it reasonable to use the stratified $k$-fold cross-validation method, which is implemented in the cross validation module of the scikit-learn library, for $k$=10.

Figure \ref{fig:TSS_features} illustrates the TSS per number of features for a prediction exactly 24 hours before the GOES X-ray flare peak time. For the red points, the features are arranged from lowest to highest F-score; for the blue points, this arrangement is reversed. In other words, Feature 1 represents the highest ranked feature, or the mean value of the horizontal field gradient, for the red points; for the blue points, Feature 1 represents the lowest ranked feature, or the mean value of the vertical field gradient. We find that any combination of 12 parameters yields the highest TSS, of approximately 0.8$\pm$0.2, whereas using fewer than 6 parameters will produce a significantly lower TSS. Based on this relatively high TSS value, we confirm that our predictive model has not overfit the data and that at least 12 of the 18 features are useful in combination. We also find that the single best feature performs marginally better on its own, producing a perverse TSS of -0.02$\pm$0.16, than the single worst feature. As such, the highest-ranked feature in our dataset does not have any predictive power on its own. Therefore, we confirm that a combination of features is necessary to distinguish between events in the positive and negative class. 

Figure \ref{fig:TSS_time} illustrates the TSS as a function of prediction time $t$, where $t$ is the number of hours before the GOES X-ray flare peak time, using all 18 features for every $t$. In this case, we predict whether or not a flaring active region will produce a CME for a $t$ that ranges from 4 to 48 hours in 4 hour intervals. We find that a $t$ of 28 to 24 hours will produce a marginally higher TSS than a $t$ of 48 or 4 hours, though the lowest TSS value is within the error for the highest one.

From Figures \ref{fig:TSS_features} and \ref{fig:TSS_time}, we find that our prediction algorithm exhibits a fairly high performance, with a TSS equal to $\sim$0.8$\pm$0.2, yet there are no other, similar studies with which to compare this value. However, since our positive class has only 56 examples, the TSS has a high error and may not generalize well to other studies. 

\section{Conclusion}
\label{section:conclude}

In this study, we [1] used features derived from maps of photospheric vector magnetic field data taken by the HMI instrument aboard SDO to forecast whether an active region that produces an M1.0-class flare or higher will also produce a CME, and [2] determined which features distinguish flaring active regions that produce CMEs from those that do not. Prior to this one, no extensive study has used physically meaningful features of active regions to distinguish between these two populations. Though we sampled from a database of 3023 active regions, we found only 56 events that satisfied the criteria defined for the positive class and 364 events for the negative one. This is due to the uncharacteristically quiet nature of Solar Cycle 24, during which time the Sun produced fewer flares and CMEs than usual. 

Through a feature selection process, we find that all the relatively high-performing features are independent of system size. In other words: as an active region increases in size, these parameters do not necessarily increase in value. Parameters that obey this property have been dubbed as {\it intensive} by \citet{welsch09}; they call the opposite case, where the value of the parameter increases with system size, {\it extensive}. Intensive parameters are generally comprised of population means, whereas extensive ones are generally comprised of population sums. 

\citet{welsch09} argue that the distinction between extensive and intensive properties is relevant to CME processes. On one hand, many numerical models show that instabilities in small-scale structures, like current sheets, form the trigger mechanism for CMEs \citep{chen11}. Changes on such a scale would be captured by an intensive parameter. On the other hand, large-scale non-potential magnetic structures, like filaments and sigmoids, have been observationally linked to CMEs \citep{webb12}. Changes in such features would be captured by an extensive parameter. 

We find that the 6 highest-ranking features, which are defined by the univariate feature selection algorithm as those that best distinguish between the positive and negative class, are all intensive in nature. Further, we find that there is a significant difference in TSS between using this group of 6 features versus using the 6 lowest-ranking features. This is illustrated in Figure \ref{fig:TSS_features}. The TSS for the 6 highest-ranking features is 0.68$\pm$0.10, whereas the TSS for the 6 lowest-ranking features is 0.19$\pm$0.19. We therefore conclude that intensive features are a better choice for predicting CMEs. 

In their observational study, \citet{sun15} also found that intensive parameters show the greatest change between flaring active regions that are CME-productive and those that are not. However, they found that the configuration of the overlying field may also play a strong role in CME productivity. As such, they suggest developing features that quantify both the relative change in intensive parameters and the topology of the overlying field. \citet{torok05} found similar results using numerical models: changes in the overlying field as a function of height strongly affects whether an eruption will lead to a confined flare or CME. Since we do not consider the magnetic field at any height above the photosphere in our study, we surmise that we lack some of the information necessary to predict CMEs.

It turns out that there is not much added value in using more than 6 features: the maximum TSS we obtain is within the error for the TSS obtained using 6 features. We compute Pearson correlation coefficients between each pair of features. While some features have almost no correlation (the correlation coefficient for the two top-ranking features in the 24-hour case is 0.06), we find that several features contain nearly identical information. For example, both of the shear-related features (ranked 6 and 7, respectively, in the 24-hour case) have a correlation coefficient of 0.99, and the total and vertical field gradients (ranked 4 and 18, respectively, in the 24-hour case) have a correlation coefficient of 0.95. Only one of each of these pairs retains a spot in the top 6 parameters. However, the second and third ranked parameters in the 24-hour case are also strongly correlated (0.98), yet both retain a spot in the top 6 parameters. This is because a high feature correlation does not indicate an absence of feature complementarity \citep {guyon03}. Rather, it appears that there is a limited amount of information contained in the photospheric magnetic field and the 6 most highly-ranked features capture most of this information. As such, we conclude that only a handful of features are necessary to obtain a relatively high TSS and that a combination of such features contains more predictive capacity than any single feature alone.

We also find that all the relatively high-performing features characterize the degree to which the magnetic field deviates from a potential configuration. We find that the mean gradient of the horizontal field holds the most predictive power of all the assessed features. \citet{sun15} found that the inclination of the horizontal field plays a strong role in whether a flaring active region produces a CME. They learned that NOAA active region 12192, which hosted the largest sunspot group in the last 24 years and produced several X-class flares yet no CMEs, contained a relatively weak horizontal field compared with other CME-producing active regions; further, active regions that exhibit a steep decrease in the magnitude of the horizontal component of the magnetic field as a function of height may be more likely to produce a CME than those that decrease much slower with height. We also find that the twist parameter, $\alpha$, is relatively high-performing. \citet{falconer02} and \citet{falconer06} found that the magnetic twist is correlated with CME productivity. They also suggest that combined features, such as $\alpha$ combined with active region size, may have an even stronger correlation with CME productivity.

When analyzing high-performing features, it is important investigate whether or not they scale with flare size. Features that do may be useful in a purely predictive sense, but they are simply a proxy of the amount of energy that is available in an active region system \citep{kahler82}. Features that do not scale with flare size, however, can provide an additional relationship with CME productivity. We find that flare class shows little difference in its distribution between the positive and negative class; further, we find no correlation between flare class and any of the SHARP features. Therefore, we conclude that the SHARP features are not simply measuring how much energy is available in an active region system but rather highlighting a physically different relationship with CME productivity.

From our study, it is unclear that the physical mechanism responsible for triggering a CME has a strong photospheric signature. Perhaps our features are too noisy; \citet{sun15} showed that local features, which parameterize the vector magnetic field within the core of an active region, show a stronger pre-CME signature than the global features we use, which parameterize the vector magnetic field within the entire active region. Or perhaps it is a better idea to also study coronal features to capture such a signature. \citet{canfield99} clearly showed that the presence of a sigmoidal shape in the X-ray corona that is simultaneous with a delta-class sunspot region on the photosphere is likely to be associated with eruptive phenomena. As such, a natural extension of this work would be to include coronal features of active regions in addition to photospheric ones.

\acknowledgments
This work was supported by NASA Grant NAS5-02139 (HMI). The data used here are courtesy of NASA/SDO and the HMI science team, as well as the GOES, STEREO, and SoHO teams. The codes used here are courtesy of the SunPy and scikit-learn teams. The authors would like to thank S\'{e}bastien Couvidat, Xudong Sun, Sanjiv Tiwari, and Yihua Zheng for their guidance.


\end{document}